\documentclass[vecphys]{svmult_pgk}		% vectors in bold face
\usepackage{cite}
\usepackage{makeidx}         % allows index generation
\usepackage{graphicx}        % standard LaTeX graphics tool
\usepackage{epsfig}          % another way to include figures
\usepackage{multicol}        % used for the two-column index
\usepackage[bottom]{footmisc}% places footnotes at page bottom

\makeindex             % used for the subject index
\begin{document}
%%%%%%%%%%%%%%%%%%%%%%%%%%%%%%%%%%%%%%%%%%%%%%%%%%%%%%%%%%%%%%%%%%%%%

% Sample title
\title*{Experimental Results Related to Discrete Nonlinear Schr\"odinger Equations}
\titlerunning{Experimental Results Related to DNLS}
\author{
Mason A. Porter\inst{1}
}

\institute{
Oxford Centre for Industrial and Applied Mathematics, Mathematical Institute, University of Oxford \texttt{porterm@maths.ox.ac.uk}
}
\maketitle

%%%%\section*{abstract}
%In this chapter, we discuss experiments that realize the discrete nonlinear Schr\"odinger (DNLS) equations.  The relevance of such descriptions arises from the competition of three common features: nonlinearity, dispersion, and a medium to large level of (periodic, quasiperiodic, or random) discreteness in space.  DNLS equations have been especially prevalent in atomic and molecular physics in the study of Bose-Einstein condensates in optical lattices or superlattices; and in nonlinear optics in the description of pulse propagation in waveguide arrays and photorefractive crystals.  New experiments in both nonlinear optics and Bose-Einstein condensation provide new challenges for DNLS models, and DNLS and related equations have also recently been used to make important predictions in novel physical settings such as the study of composite metamaterials and arrays of superconducting devices. 

\section{Introduction}\label{sec8.1_intro}

Discrete nonlinear Schr\"odinger (DNLS) equations can be used to model numerous phenomena in atomic, molecular, and optical physics.  The general feature of these various settings that leads to the relevance of DNLS models is a competition between nonlinearity, dispersion, and spatial discreteness (which can be periodic, quasiperiodic, or random).  In three-dimensions (3D), the DNLS with cubic nonlinearity is written in normalized form as
\begin{equation}\label{sec8.1_dnls}
	i\dot{u}_{l,m,n} = - \epsilon \Delta u_{l,m,n} \pm |u_{l,m,n}|^2 u_{l,m,n} + V_{l,m,n}(t)u_{l,m,n}\,,
\end{equation}
where $\Delta$ is the discrete Laplacian; $\epsilon$ is a coupling constant; $u_{l,m,n}$ is the value of the field at site $(l,m,n)$; and $V_{l,m,n}(t)$ is the value of the external potential at that site.  In Eq.~(\ref{sec8.1_dnls}), a $+$ sign represents the defocusing case and a $-$ sign represents the focusing one.  Both of these situations have been discussed extensively throughout this book.

The study of DNLS equations dates back to theoretical work on biophysics in the early 1970s \cite{sec8.1_davydov73}.  In the late 1980s, this early research motivated extensive analysis of such equations for the purpose of modeling the dynamics of pulses in optical waveguide arrays \cite{sec8.1_chris88}.  One decade later, experiments using fabricated Aluminum gallium arsenide (AlGaAs) waveguide arrays \cite{sec8.1_eisen} stimulated a huge amount of subsequent research, including experimental investigations of phenomena such as discrete diffraction, Peierls barriers, diffraction management \cite{sec8.1_moran,sec8.1_eisen00}, gap solitons \cite{sec8.1_mandel}, and more \cite{sec8.1_moran01}.  As was first suggested theoretically in \cite{sec8.1_efrem02} and realized experimentally in  \cite{sec8.1_fleischer03,sec8.1_fleischer03prl,sec8.1_neshev03,sec8.1_martin}, DNLS equations also accurately predict the existence and stability properties of nonlinear localized waves in optically-induced lattices in photorefractive media such as Strontium Barium Niobate (SBN).  Because of this success, research in this arena has exploded; structures such as dipoles \cite{sec8.1_yang04}, quadrupoles \cite{sec8.1_yang04b}, multiphase patterns (including soliton necklaces and stripes) \cite{sec8.1_yang05,sec8.1_neshev04}, discrete vortices \cite{sec8.1_neshev04prl,sec8.1_fleischer04}, and rotary solitons \cite{sec8.1_wang06} have now been theoretically predicted and experimentally obtained in lattices induced with a self-focusing nonlinearity.  As discussed in Ref.~\cite{sec8.1_tang07} (and references therein), self-defocusing realizations have also been obtained.  These allow the construction of dipole-like gap solitons, etc.

DNLS equations have also been prominent in investigations of Bose-Einstein condensates (BECs) in optical lattice (OL) potentials, which can be produced by counter-propagating laser beams along one, 
two, or three directions \cite{sec8.1_oberrev}.  This field has also experienced enormous growth in the last ten years; major experimental results that have been studied using DNLS equations include modulational (``dynamical") instabilities \cite{sec8.1_smer02,sec8.1_cata}, gap soliton dynamics \cite{sec8.1_markus2}, Bloch oscillations and Landau-Zener tunnelling \cite{sec8.1_anderson}, and the production of period-doubled solutions \cite{sec8.1_chu}.

\section{Optics}\label{sec8.1_optics}

For many decades, optics has provided one of the traditional testbeds for investigations of nonlinear wave propagation \cite{sec8.1_photon}.  For example, the (continuous) nonlinear Schr\"odinger (NLS) equation provides a dispersive envelope wave model for describing the electric field in optical fibers.  In the presence of a spatially-discrete external potential (such as a periodic potential), one can often reduce the continuous NLS to the discrete NLS.  In this section, we will consider some appropriate situations that arise in optical waveguide arrays and photorefractive crystals.  Many of these optical phenomena have almost exact analogs in both solid state and atomic physics \cite{sec8.1_scott}.

\subsection{Optical Waveguide Arrays}

%\begin{figure}
%\center
%\includegraphics[width=10cm]{yaron.eps}
%\caption{Periodic array of single-mode waveguides.  (Top) Image produced by a scanning electron microscope.  (Bottom) Illustration of layered structure.  (Image from Ref.~\cite{sec8.1_suk03}.) }
%\label{sec8.1_waveguide}       % Give a unique label
%\end{figure}

\begin{figure}
\center
\includegraphics[width=5cm]{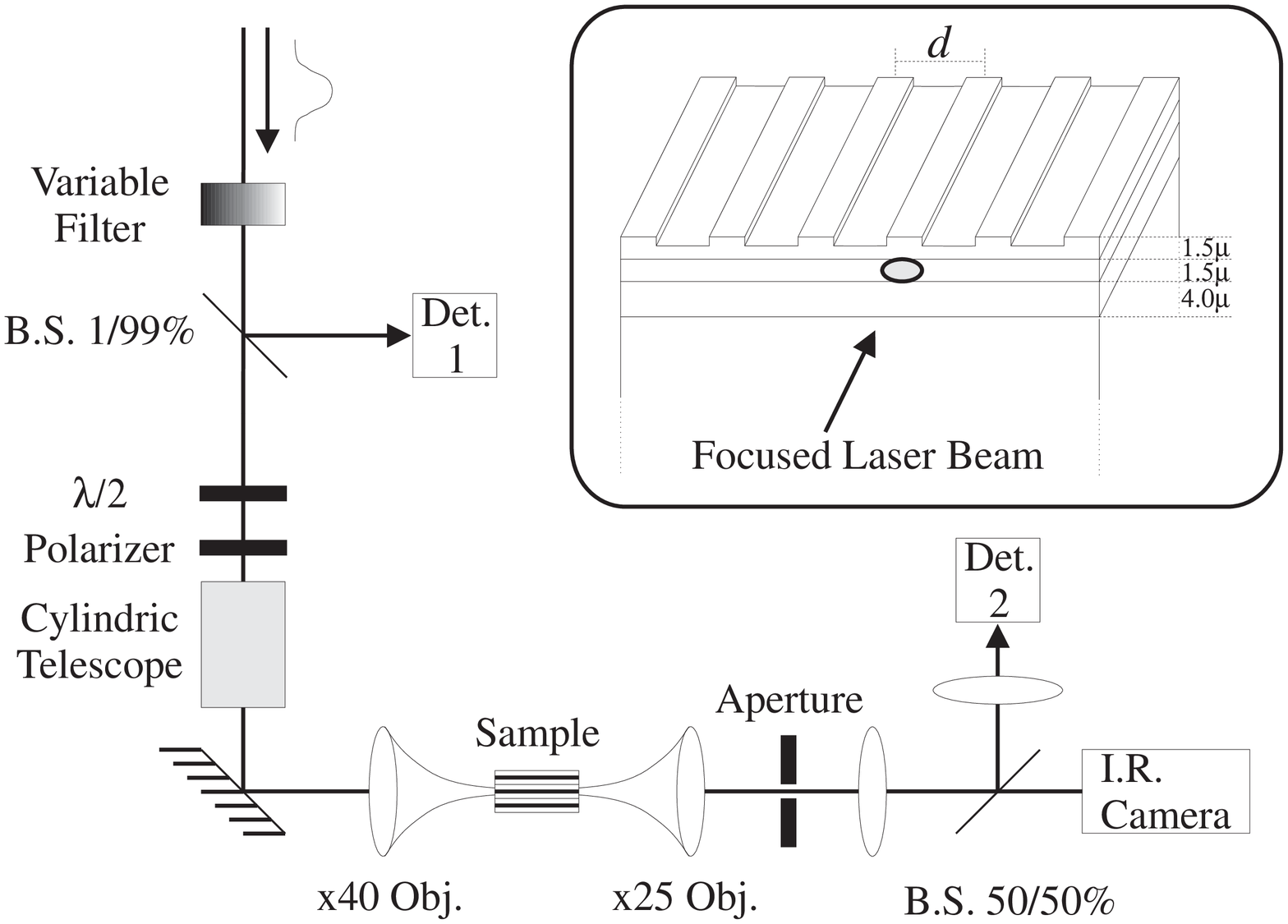}
\hspace{.2cm}
\includegraphics[width=5cm]{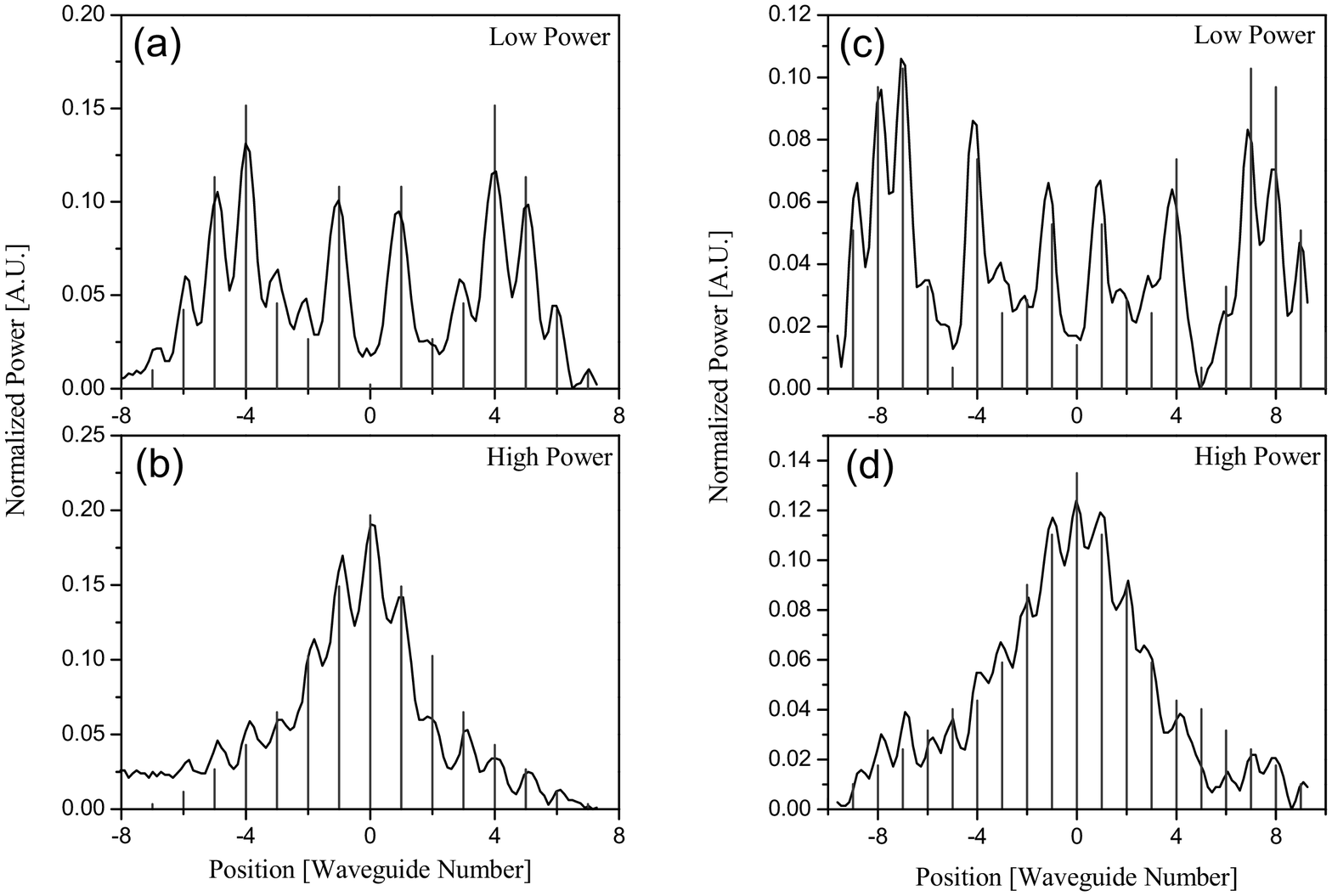}
\caption{(Left) Experimental setup for the waveguide array experiments reported in Ref.~\cite{sec8.1_eisen}.  (Right) Low power (diffraction) versus high power experiments.  The latter result in discrete spatial solitons.  (Images from Ref.~\cite{sec8.1_eisen}.)}
\label{sec8.1_waveguide}       % Give a unique label
\end{figure}

\begin{figure}
\center
\includegraphics[width=10cm]{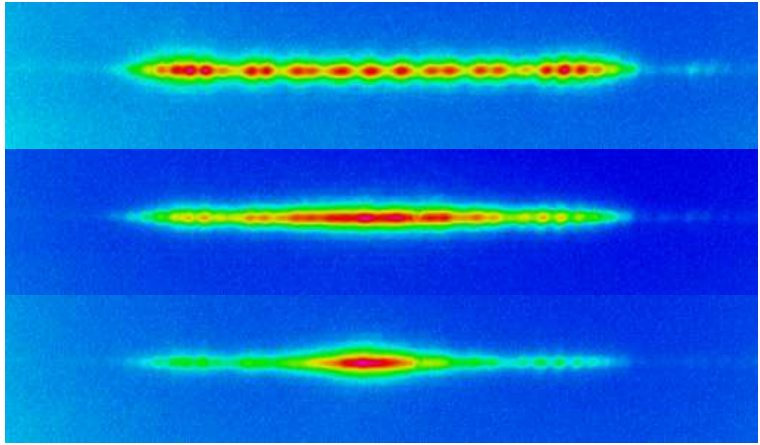}
\caption{Experimental observation of a soliton (bottom panel) at the output facet of a waveguide array.  The peak powers are 70 W (top), 320 W (center), and 500 W (bottom).  (Image from Ref.~\cite{sec8.1_eisen}.)}
\label{sec8.1_soliton}
\end{figure}

An optical waveguide is a physical structure that guides electromagnetic waves in the optical spectrum.   Early proposals in nonlinear optics suggested that light beams can trap themselves by creating their own waveguide through the nonlinear Kerr effect \cite{sec8.1_agrawal}.  Waveguides confine the diffraction, allowing spatial solitons to exist.  In the late 1990s, Eisenberg et al. showed experimentally that a similar phenomenon (namely, discrete spatial solitons) can occur in a coupled array of identical waveguides \cite{sec8.1_eisen}.  One injects low-intensity light into one waveguide (or a small number of neighboring ones); this causes an ever increasing number of waveguides to couple as it propagates, analogously to what occurs in continuous media.  If the light has high intensity, the Kerr effect changes the refractive index of the input waveguides, effectively decoupling them from the rest of the array.  That is, certain light distributions propagate with a fixed spatial profile in a limited number of waveguides; see Figs.~\ref{sec8.1_waveguide} and \ref{sec8.1_soliton}.

The standard theoretical approach used to derive a DNLS equation in the context of one-dimensional (1D) waveguide arrays is to decompose the total field (describing the envelope amplitude) into a sum of weakly coupled fundamental modes that are excited in each individual waveguide \cite{sec8.1_chris88,sec8.1_suk03}.  If one supposes that each waveguide is only coupled to its nearest neighbors, this approach amounts to the {\it tight-binding approximation} of solid state physics.  For lossless waveguides with a Kerr (cubic) nonlinearity, one obtains the DNLS 
\begin{equation}\label{sec8.1_chris}
	i\frac{dE_n}{dz} + \beta E_n + c(E_{n+1} + E_{n-1}) + \lambda |E_n|^2E_n + \mu (|E_{n+1}|^2 + |E_{n-1}|^2)E_n = 0\,,
\end{equation}
where $E_n$ is the mode amplitude of the $n$th waveguide, $z$ is the propagation direction, $\beta$ is the field propagation constant of each waveguide, $c$ is a coupling coefficient, and $\lambda$ and $\mu$ are positive constants that respectively determine the strengths of the self-phase and cross-phase modulations experienced by each waveguide.  The quantity $\lambda$ is proportional to the optical angular frequency and Kerr nonlinearity coefficient, and is inversely proportional to the effective area of the waveguide modes \cite{sec8.1_suk03}.  In most situations, the self-phase modulation dominates the cross-phase modulation (which arises from the nonlinear overlap of adjacent modes), so that $\mu \ll \lambda$.  This allows one to set $\mu = 0$ and yields the model that Davydov employed for $\alpha$-spiral protein molecules \cite{sec8.1_davydov73}.  The transformation $E_n(z) = \Phi_n(z)\exp[i(2c + \beta)z]$ and a rescaling then gives a 1D version of Eq.~(\ref{sec8.1_dnls}).

The impact of Ref.~\cite{sec8.1_eisen} was immediate and powerful, as numerous subsequent experiments reported very interesting phenomena.  For example, this setting provided the first experimental demonstration of the Peierls-Nabarro (PN) potential in a macroscopic system \cite{sec8.1_moran}, thereby explaining the strong localization observed for high-intensity light in the original experiments \cite{sec8.1_eisen}.  That is, the PN potential describes the energy barrier between the (stable) solitons that are centered on a waveguide and propagate along the waveguide direction and the (unstable) ones that are centered symmetrically between waveguides and tend to shift away from the waveguide direction.  Eisenberg et al. \cite{sec8.1_eisen00} have also exploited diffraction management (which is analogous to the dispersion management ubiquitously employed in the study of temporal solitons \cite{sec8.1_borisbook}) to produce structures with designed (reduced, canceled, or reversed) diffraction properties.  The ability to engineer the diffraction properties has paved the way for new possibilities (not accessible in bulk media) for controlling light flow.  For example, using two-dimensional (2D) waveguide networks, discrete solitons can travel along essentially arbitrarily curves and be routed to any destination \cite{sec8.1_christ03}.  This may prove extremely helpful in the construction of photonic switching architectures.

A waveguide array with linearly increasing effective refractive index, which can be induced using electro- or thermo-optical effects, has also been used to demonstrate Bloch oscillations (periodic recurrences) in which the initial distribution is recovered after one oscillation period \cite{sec8.1_moran99}.  A single-waveguide excitation spreads over the entire array before refocusing into the initial guide.  More recently, Morandotti et al. investigated the interactions of discrete solitons with structural defects produced by modifying the spacing of one pair of waveguides in an otherwise uniform array \cite{sec8.1_moran03}.  This can be used to adjust the PN potential.  It has also been demonstrated experimentally that even a binary array is sufficient to generate discrete gap solitons, which can then be steered via inter-band momentum exchange \cite{sec8.1_moran04}.  From a nonlinear dynamics perspective, an especially exciting result is the experimental observation of discrete modulational instabilities \cite{sec8.1_meier}.  Using an AlGaAs waveguide array with a self-focusing Kerr nonlinearity, Meier et al. found that such an instability occurs when the initial spatial Bloch momentum vector is within the normal diffraction region of the Brillouin zone.  (It is absent even at very high power levels in the anomolous diffraction regime.)  More recent experimental observations include discrete spatial gap \cite{sec8.1_chenlith} and dark \cite{sec8.1_smirnov06} solitons in photovoltaic lithium niobate (LiNbO$_3$) waveguide arrays, evidence for the spontaneous formation of discrete $X$ waves in AlGaAs waveguide arrays \cite{sec8.1_lahini07}, and an analog of Anderson localization (which occurs in solid state physics when an electron in a crystal becomes immobile in a disordered lattice) \cite{sec8.1_lahini08}.

\subsection{Photorefractive Crystals}\label{sec8.1_photo}

\begin{figure}
\center
\includegraphics[width=5cm]{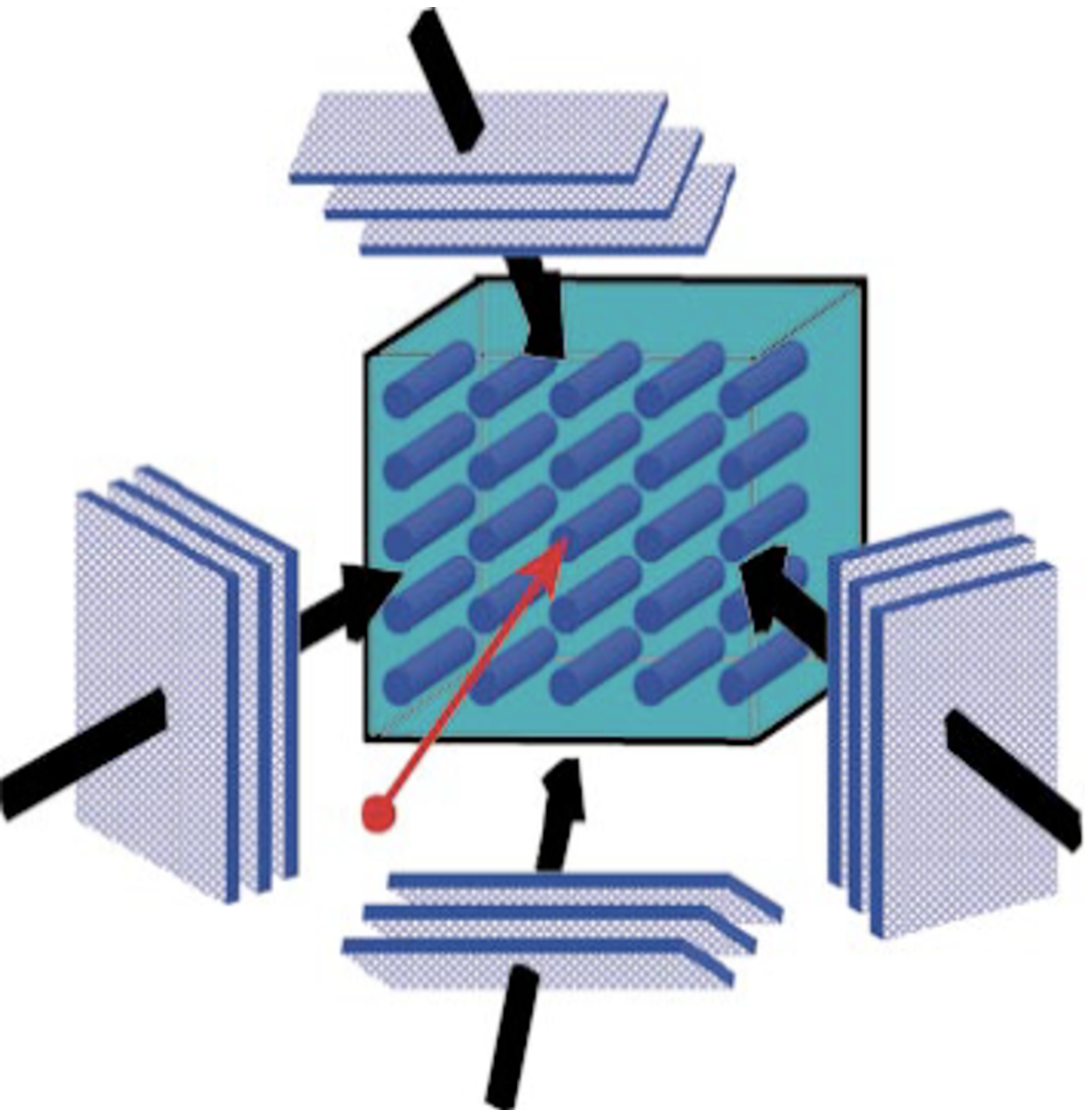}
\hspace{1cm}
\includegraphics[width=5cm]{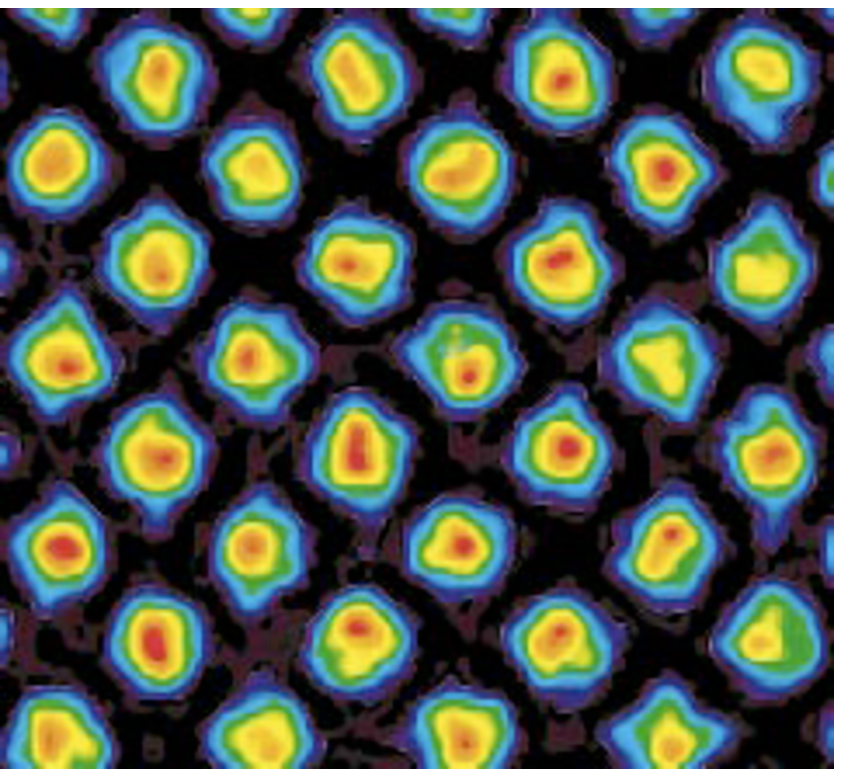}
\caption{(Left) Diagram of experimental setup for the creation of a photorefractive crystal lattice with electro-optic anisotropy.  (Right) Typical observation of the lattice at the terminal face of the crystal.  Each waveguide has a diameter of about $7 \mu m$ and is about $11 \mu m$ away from its nearest neighbors.  (Images from Ref.~\cite{sec8.1_fleischer03}.)  
}
\label{sec8.1_photo_pic}
\end{figure}

Photorefractive crystals can be used to construct 2D periodic lattices via plane wave interference by employing a technique known as optical induction.  This method, which was developed theoretically in Ref.~\cite{sec8.1_efrem02} and subsequently demonstrated experimentally for 1D discrete solitons in \cite{sec8.1_fleischer03prl} and bright 2D solitons in \cite{sec8.1_fleischer03}, has become a very important playground for investigations of nonlinear waves in optics \cite{sec8.1_fleischer05oe}.  One obtains a periodic lattice in real time through the interference of two or more plane waves in a photosensitive material (see Fig.~\ref{sec8.1_photo_pic}).  One then launches a probe beam, which experiences discrete diffraction (the optical equivalent of quantum tunnelling in a periodic potential) and can form a discrete soliton provided the nonlinearity is sufficiently large.  The model for photorefractive crystals is a continuous NLS equation with saturable nonlinearity \cite{sec8.1_fleischer03},
\begin{equation}
	iU_z + U_{xx} + U_{yy} - \frac{E_0}{1 + I_l + |U|^2}U = 0\,,
\end{equation}
where $z$ is the propagation distance, $(x,y)$ are transverse coordinates, $U$ is the slowly-varying amplitude of the probe beam (normalized by the dark irradiance of the crystal), $E_0$ is the applied dc field, and $I_l$ is a lattice intensity function.  For a square lattice, $I_l = I_0\sin^2\{(x+y)/\sqrt{2}\}\sin^2\{(x-y)/\sqrt{2}\}$, where $I_0$ is the lattice's peak intensity.  DNLS equations have been enormously insightful in providing corroborations between theoretical predictions and experimental observations (see, in particular, the investigations of discrete vortices in Refs.~\cite{sec8.1_neshev04prl,sec8.1_fleischer04,sec8.1_chen05opt}), although they do not provide a prototypical model in this setting the way they do with waveguide arrays.  

For optical induction to work, it is essential that the interfering waves are unaffected by the nonlinearity (to ensure that the ``waveguides" are as uniform as possible) but that the probe (soliton-forming) beam experiences a significant nonlinearity.  This can be achieved by using a photorefractive material with a strong electro-optic anisotropy.  In such materials, coherent rays interfere with each other and form a spatially-varying pattern of illumination (because the local index of refraction is modified, via the electro-optic effect, by spatial variations of the light intensity).  This causes ordinary polarized plane waves to propagate almost linearly (i.e., with practically no diffraction) and extraordinary polarized waves to propagate in a highly nonlinear fashion.  The material of choice in the initial experiments of Ref.~\cite{sec8.1_fleischer03} was the (extremely anisotropic) SBN:75 crystal.
 
The theoretical prediction and subsequent experimental demonstration of 2D discrete optical solitons has led to the construction and analysis of entirely new families of discrete solitons \cite{sec8.1_efrem02,sec8.1_fleischer03}.  As has been discussed throughout this book, 
%({appropriate section references to be inserted here}), 
the extra dimension allows much more intricate nonlinear dynamics to occur than is possible in the 1D waveguides discussed above.  Early experiments demonstrated novel self-trapping effects such as the excitation of odd and even nonlinear localized states \cite{sec8.1_neshev03}.  They also showed that photorefractive crystals can be used to produce index gratings that are more controllable than those in fabricated waveguide arrays.  

Various researchers have since exploited the flexibility of photorefractive crystals to create interesting, robust 2D structures that have the potential to be used as carriers and/or conduits for data transmission and processing in the setting of all-optical communication schemes (see Refs.~\cite{sec8.1_fleischer05oe,sec8.1_chen05opt} and references therein).   In the future, 1D arrays in 2D environments might be used for multidimensional waveguide junctions, which has the potential to yield discrete soliton routing and network applications.  For example, Martin et al. observed soliton-induced dislocations and deformations in photonic lattices created by partially incoherent light \cite{sec8.1_martin}.  By exploiting the photorefractive nonlinearity's anisotropy, they were able to create optical structures analogous to polarons\footnote{A polaron is a quasiparticle composed of a conducting electron and an induced polarization field that moves with the electron.} from solid state physics \cite{sec8.1_ashcroft}.  An ever-larger array of structures has been predicted and experimentally obtained in lattices induced with a self-focusing nonlinearity.  For example, Yang et al. demonstrated discrete dipole (two-hump) \cite{sec8.1_yang04} and quadrupole (four-hump) solitons \cite{sec8.1_yang04b} both experimentally and theoretically.  They also showed that both dipole and quadrupole solitons are stable in a large region of parameter space when their humps are out of phase with each other.  (The stable quadrupole solitons were square-shaped; adjacent humps had a phase difference of $\pi$, so that diagonal humps had the same phase.)  Structures that have been observed in experiments in the self-defocusing case include dipole-like gap solitons \cite{sec8.1_tang07} and gap-soliton vortices \cite{sec8.1_panospersonal}.

More complicated soliton structures have also been observed experimentally.  For example, appropriately launching a high-order vortex beam (with, say, topological charge $m = 4$) into a photonic lattice can produce a stationary necklace of solitons \cite{sec8.1_yang05}.  Stripes of bright \cite{sec8.1_neshev04} and gap \cite{sec8.1_wang07} solitons have also been created \cite{sec8.1_neshev04}, providing an interesting connection with several other pattern-forming systems \cite{sec8.1_cross}.  Another very fruitful area has been the construction of both off-site and on-site discrete vortices \cite{sec8.1_neshev04prl,sec8.1_fleischer04,sec8.1_chen05opt}.  Recent experimental observations in this direction have included self-trapping and charge-flipping of double-charged optical vortices (which lead to the formation of rotating quasi-vortex solitons) \cite{sec8.1_anna06}.  Discrete rotary solitons \cite{sec8.1_wang06} and discrete random-phase solitons \cite{sec8.1_cohen05} have also been observed.  In fact, the results of Ref.~\cite{sec8.1_cohen05} are reminiscent of the Fermi-Pasta-Ulam (FPU) numerical experiments \cite{sec8.1_focus}, as an initially homogeneous distribution in momentum space evolved into a steady-state multi-humped soliton power spectrum.

\begin{figure}
\center
\includegraphics[width=10cm]{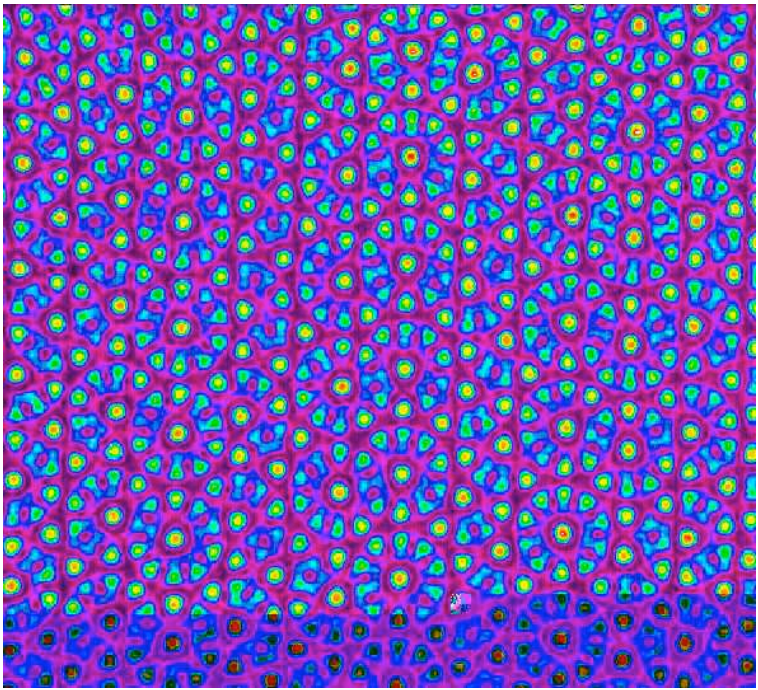}
\caption{Experimental image of the decagonal field-intensity pattern in an optically-induced nonlinear photonic quasicrystal.  (Image from Ref.~\cite{sec8.1_lif06}.)}
% {\bf map: possibly change to black and white} 
\label{sec8.1_crystal}
\end{figure}

The investigation of photorefractive crystals continues to produce experimental breakthroughs, offering ever more connections to solid state physics.  One particularly exciting experiment was the observation of dispersive shock waves \cite{sec8.1_shock07}.  Another fascinating result was the observation of an analog of Anderson localization in disordered 2D photonic lattices \cite{sec8.1_fish07}.  In this context, the transverse localization of light is caused by random fluctuations.  Wave, defect, and phason dynamics (including discrete diffraction and discrete solitons) have recently been investigated experimentally in optically-induced nonlinear photonic {\it quasicrystals} \cite{sec8.1_lif06,sec8.1_lif07} (see Fig.~\ref{sec8.1_crystal}), whose theoretical investigation provides one of the outstanding challenges for DNLS models.

\section{Bose-Einstein Condensation}\label{sec8.1_bec}

\begin{figure}
\center
\includegraphics[width=10cm]{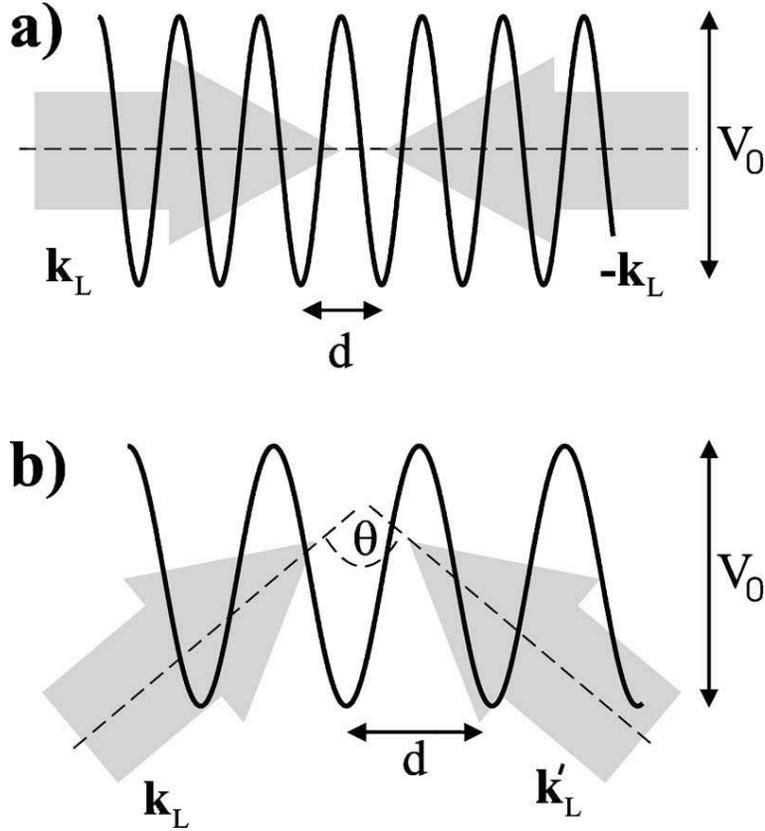}
\caption{Diagram of the creation of a 1D optical lattice potential using (a) counterpropagating laser beams and (b) beams intersecting at angle $\theta$.  The quantities ${\bf k}_L$ and ${\bf k}'_L$ denote the wave vectors of the beams.  The lattice period is given by the distance $d$ between consecutive maxima of light intensity in the interference pattern.  (Images from Ref.~\cite{sec8.1_oberrev}.)} 
\label{sec8.1_ober1}
\end{figure}

\begin{figure}
\center
\includegraphics[width=10cm]{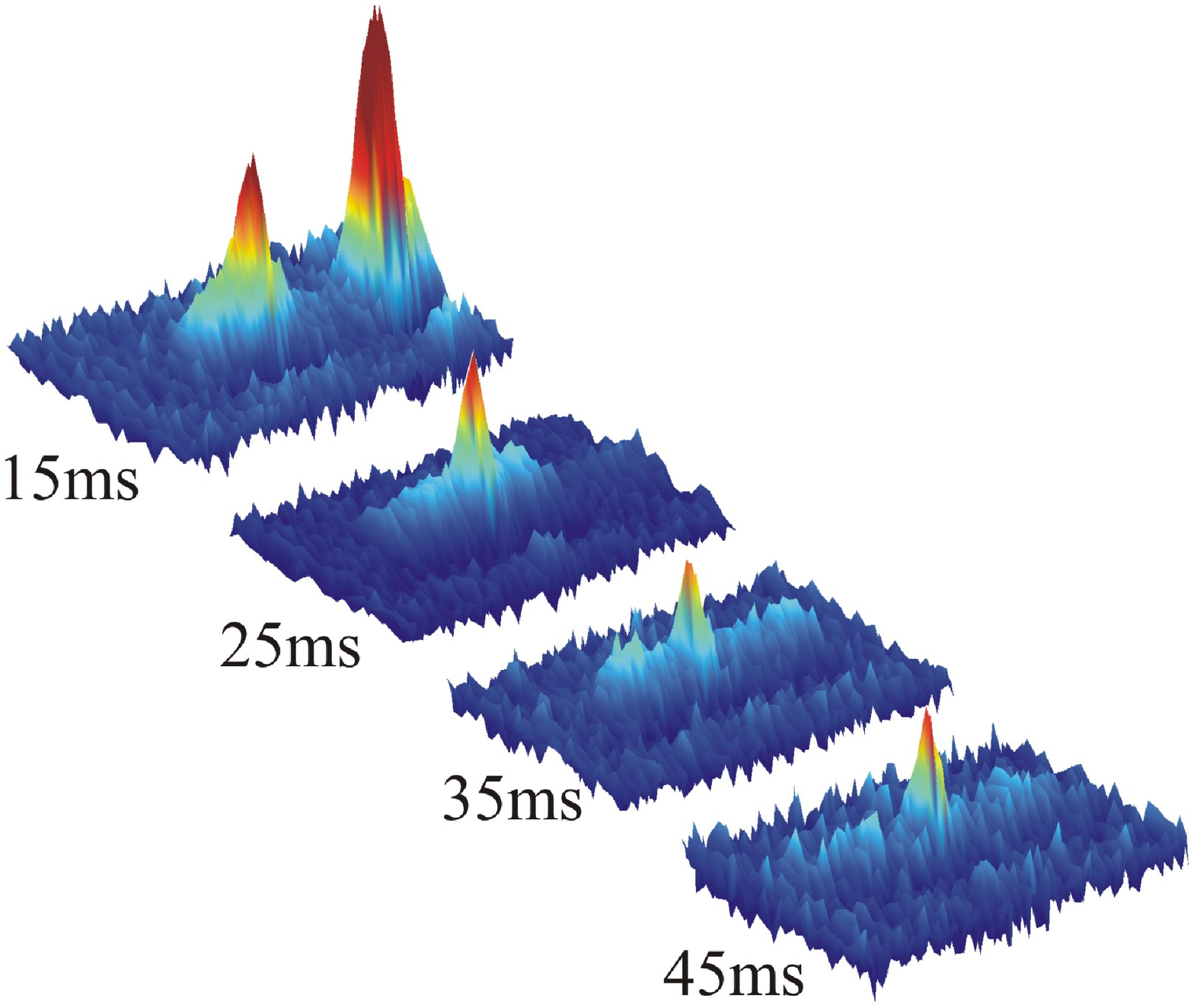}
\caption{Experimental demonstration of a bright gap soliton (the shading shows the atomic density).  
%The color shows the atomic density, with low values in blue and high values in red.  
A small, stable peak forms after about 25 ms.  (Image from Ref.~\cite{sec8.1_markus2}.)
}
\label{sec8.1_ober2}
\end{figure}

At sufficiently low temperature, bosonic particles in a dilute 3D gas occupy the same quantum (ground) state, forming a Bose-Einstein condensate (BEC) \cite{sec8.1_pethick,sec8.1_stringari,sec8.1_ketter,sec8.1_edwards}. Seventy years 
after they were first predicted theoretically, dilute (i.e., weakly interacting) BECs were finally observed experimentally in 1995 in vapors of rubidium and sodium \cite{sec8.1_becrub,sec8.1_becna}. In these experiments, atoms were loaded into magnetic traps and evaporatively cooled to temperatures well below a microkelvin. To record the properties of the BEC, the confining trap was then switched off, and the expanding gas was optically imaged \cite{sec8.1_stringari}. A sharp peak in the velocity distribution was observed below a critical temperature $T_c$, indicating that condensation had occurred.  

If the temperature is well below $T_c$, then considering only two-body, mean-field interactions, the BEC dynamics is modeled using the 3D Gross-Pitaevskii (GP) equation (i.e., the continuous cubic NLS equation),
\begin{equation}
  	i\hbar {\Psi }_{t}=\left( -\frac{\hbar ^{2}\nabla ^{2}}{2m}+g_{0}|\Psi|^{2}+{\mathcal{V}}(\vec{r})\right) \Psi\,,  \label{sec8.1_GPE}
\end{equation}
where $\Psi =\Psi (\vec{r},t)$ is the condensate wave function (order parameter)
normalized to the number of atoms, ${\mathcal{V}}(\vec{r})$ is the external potential, and the
effective self-interaction parameter is $\tilde{g}=[4\pi\hbar ^{2}a/m][1+O(\zeta ^{2})]$, where $a$ is the two-body scattering length and $\zeta \equiv \sqrt{|\Psi |^{2}|a|^{3}}$ is the dilute-gas parameter \cite{sec8.1_stringari,sec8.1_kohler,sec8.1_baiz}.  The cubic nonlinearity arises from the nearly perfect contact (delta-function) interaction between particles.

In a quasi-1D (``cigar-shaped") BEC, the transverse dimensions are about equal to the healing length, and the longitudinal dimension is much larger than the transverse ones.  One can then average (\ref{sec8.1_GPE}) in the transverse plane to obtain the 1D GP equation \cite{sec8.1_salasnich,sec8.1_stringari},
\begin{equation}
	      i\hbar u_t = -[\hbar^2/(2m)] u_{xx} + g|u|^2 u + V(x) u \,, \label{sec8.1_nls3}
\end{equation}
where $u$, $g$, and $V$ are, respectively, the rescaled 1D wave function, interaction parameter, and external trapping potential.  The interatomic interactions in BECs are determined by the sign of $g$: they are repulsive (producing a defocusing nonlinearity) when $g > 0$ and attractive (producing a focusing nonlinearity) when $g < 0$.

BECs can be loaded into optical lattice (OL) potentials (or superlattices, which are small-scale lattices subjected to a large-scale modulation), which are created experimentally as interference patterns of laser beams.  Consider two identical laser beams with parallel polarization and equal peak intensities, and counterpropagate them as in Fig.~\ref{sec8.1_ober1}a so that their cross sections overlap completely.  The two beams create an interference pattern with period $d = \lambda_L/2$ (half of the optical wavelength) equal to the distance between consecutive maxima of the resulting light intensity.  The potential experienced by atoms in the BEC is then \cite{sec8.1_oberrev}
\begin{equation}
	V(x) = V_0\cos^2(\pi x/d)\,,
\end{equation}
where $V_0$ is the lattice depth.  See Ref.~\cite{sec8.1_oberrev} for numerous additional details.

BECs were first successfully placed in OLs in 1998 \cite{sec8.1_anderson}, and numerous labs worldwide now have the capability to do so.  Experimental and theoretical investigations of BECs in OLs (and related potentials) have developed into one of the most important subdisciplines of BEC investigations \cite{sec8.1_pethick,sec8.1_oberrev,sec8.1_fpubec}.  We focus here on results that can be modeled using a DNLS framework; see the reviews \cite{sec8.1_oberrev,sec8.1_review08} for discussions of and references to myriad other outstanding experiments.  
%(Reference \cite{sec8.1_review08} gives another recent review of research on BECs in OL potentials.)  
The first big experimental result was the observation of Bloch oscillations in a repulsive BEC by Anderson and Kasevich \cite{sec8.1_anderson}.  They used a trapping potential with both an OL and a linear (gravitational) component to create a sloping periodic (``washboard") potential.  When the slope was small, wave packets remained confined in a single band and oscillated coherently at the Bloch frequency.  This effect is closely related to the ac Josephson effect in superconducting electronic systems.  For larger slopes, wave packets were able to escape their original band and transition to higher states.  Trombettoni and Smerzi analyzed the results of these experiments using a DNLS equation that they derived from the GP equation with the appropriate (OL plus gravitational) potential using the tight-binding approximation valid for moderate-amplitude potentials in which Bloch waves are strongly localized in potential wells \cite{sec8.1_smer}.  In this paper, they also showed that discrete breathers (specifically, bright gap solitons) can exist in BECs with repulsive potentials (i.e., defocusing nonlinearities).  

Other fundamental experimental work on BECs in OLs has concerned superfluid properties.  In 2001, Burger et al. treated this setting as a homogeneous superfluid with density-dependent critical velocity \cite{sec8.1_lattice}.  Cataloiotti et al. then built on this research to examine a classical transition between superfluid and Mott insulator behavor in BECs loaded into an OL superimposed on a harmonic potential \cite{sec8.1_cata}.  (The better-known quantum transition was first shown in Ref.~\cite{sec8.1_mott}.)  The BEC exhibits coherent oscillations in the ``superfluid" regime and localization in the harmonic trap in the ``insulator" regime, in which each site has many atoms of its own and is effectively its own BEC.  The transition from superfluidity to Mott insulation occurs when the condensate wave packet's initial displacement is larger than some critical value or, equivalently, when the velocity of its center of mass is larger than a critical velocity that depends on the tunnelling rate between adjacent OL sites.  These experiments confirmed the predictions of Ref.~\cite{sec8.1_smer02}, which used a DNLS approach to predict the onset of this superfluid-insulator transition via a discrete modulational (``dynamical") instability and to derive an analytical expression for the critical velocity at which it occurs.

Subsequent theoretical work with DNLS equations predicted that modulational instabilities could lead to ``period-doubled" solutions in which the BEC wavefunction's periodicity is twice that of the underlying OL \cite{sec8.1_pethick2}.  (Period-doubled wavefunctions were simultaneously constructed using a GP approach \cite{sec8.1_mapbecprl}.)  The modulational instability mechanism was exploited experimentally the next year to construct these solutions by parametrically exciting a BEC via periodic translations (shaking) of the OL potential \cite{sec8.1_chu}.  Parametric excitation of BECs promises to lead to many more interesting insights in the future.

By balancing the spatial periodicity of the OL with the nonlinearity in the DNLS, one can also construct intrinsic localized modes (discrete breathers) known as bright gap solitons, which resemble those supported by Bragg gratings in nonlinear optical systems.  In BECs, such breathers have been predicted in two situations: 
\begin{enumerate}
	\item{The small-amplitude limit in which the value of chemical potential is close to forbidden zones (``gaps'') of the underlying linear Schr\"{o}dinger equation with a periodic
potential \cite{sec8.1_kon}.}
	\item{In the tight-binding approximation, for which the continuous NLS equation with a periodic potential can be reduced to the DNLS equation \cite{sec8.1_smer02}.  (As mentioned earlier, this corresponds to the standard manifestation of the DNLS in waveguide arrays.)} 
\end{enumerate}
Recent experiments \cite{sec8.1_markus2} have confirmed the first prediction (see Fig.~\ref{sec8.1_ober2}). 

Another important development was the experimental construction of 2D and 3D OLs \cite{sec8.1_oberrev}, which as in optical systems leads to much more intricate nonlinear localized structures.  One can obtain higher-dimensional OL potentials by using additional pairs of laser beams.  The simplest way to do this is to have pairs of counterpropagating laser beams along each of two or three mutually orthogonal axes.  The interference pattern obtained with this many laser beams depends sensitively on their polarizations, relative phases, and orientations.  This allows experimentalists to construct a large variety of OL geometries in 2D and 3D.  In 2001, Greiner et al. showed that BECs can be efficiently transferred into 2D lattice potentials by adiabatically increasing the depth of the lattice \cite{sec8.1_greiner01a,sec8.1_greiner01b}.  They confined atoms to an array of narrow potential tubes, each of which was filled with a 1D quantum gas.  Around the same time, Burger et al. confined quasi-2D BECs into the lattice sites of a 1D OL potential \cite{sec8.1_burger02}.  By adding more laser beams and/or controlling their polarizations and relative phases, experimentalists can in principle create even more complicated potentials (such as quasiperiodic or Kagom\'e lattices) \cite{sec8.1_santos04}.  Very recently, there has been also been a great deal of theoretical and experimental interest in rotating optical lattices \cite{sec8.1_tung06}, which can likely be modeled using an appropriate DNLS framework to study interesting vortex dynamics.  Such  experiments and (more generally) investigations of BECs with effective fields obtained by this and other \cite{sec8.1_juze06} means promise to yield considerable insights into quantum hall physics.  

In addition to the fascinating insights into basic physics discussed above, two other major consequences of investigations of BECs in OLs and related potentials have been to bring quantum computation one (small) step closer to reality and to help bridge the gap between condensed matter physics and atomic/molecular physics.  One of the key proposed systems for constructing a quantum computer is a BEC in optical lattice and related potentials \cite{sec8.1_voll,sec8.1_ciractoday}.  This was the motivation for the experimental implementation of optical superlattice potentials \cite{sec8.1_quasibec} and its 2D egg-carton descendent, which consists of an optical lattice potential in one cardinal direction and a double-well potential in another \cite{sec8.1_ander06}.  This has led very recently to the experimental realization of a two-qubit quantum gate \cite{sec8.1_ander07}.  The second front has been advanced experimentally in the OL context by investigations of Fermi condensates in OL potentials \cite{sec8.1_mod03,sec8.1_kohl05}, Bose-Fermi mixtures in OL potentials \cite{sec8.1_ott04}, and more.  Finally, in parallel with the recent insights in optics discussed above, Anderson localization has been observed recently both for a BEC placed in a 1D waveguide with controlled disorder \cite{sec8.1_becanderson} and for a BEC in a 1D quasiperiodic OL \cite{sec8.1_becanderson2}.

\section{Summary and Outlook}\label{sec8.1_sum}

In this review, I have discussed applications in nonlinear optics and Bose-Einstein condensation in which DNLS equations have been used to explain fundamental and striking experimental results.  In optics, DNLS equations provide a prototypical model for the dynamics of discrete solitons in waveguide arrays.  The same is true for BECs in optical lattice (and related) potentials.  DNLS equations have also been used successfully to predict robust experimental features in photorefractive crystals, although they do not provide a prototypical model in this setting.

DNLS equations arise in a number of other contexts as envelope models for several types of nonlinear lattice equations (such as ones of Klein-Gordon type).  Related experiments have revealed the existence of intrinsic localized modes in these systems \cite{sec8.1_sato06}.  Relevant settings include quasi-1D antiferromagnets \cite{sec8.1_wrubel05}, micromechanical oscillator arrays \cite{sec8.1_sato06}, and electric transmission lines \cite{sec8.1_sato07}.  

DNLS equations have also been used in a variety of other settings to make interesting predictions that have not yet been verified experimentally.  For example, in composite metamaterials, Shadrivov et al. have analyzed the modulational instability of different nonlinear states and demonstrated that nonlinear metamaterials support the propagation of domain walls (kinks) that connect regions of positive and negative magnetization \cite{sec8.1_shad06}.  Very recently, dissipative discrete breathers were constructed in a model of rf superconducting quantum interference device (SQUID) arrays \cite{sec8.1_squid07}.  (Similar discrete breathers have also recently been studied theoretically in both 1D and 2D in the setting of metamaterials \cite{sec8.1_meta07}.)  This model is reminiscent of a DNLS with dissipation, except that the nonlinearity was sinusoidal rather than cubic.  A bit farther afield, 1D chains of granular materials (sometimes called ``phononic crystals") have been given increasing attention from both experimentalists and theorists in recent years \cite{sec8.1_nesterenko1,sec8.1_dar05}.  When given an initial precompression, they can exhibit optical modes that are expected to be describable as gap solitons in a nonlinear lattice model reminiscent of FPU chains.

In conclusion, DNLS equations and related models have been incredibly successful in the description of numerous experiments in nonlinear optics and Bose-Einstein condensation.  They also show considerable promise in a number of other settings, and related nonlinear lattice models are also pervasive in a huge number of applications.  To borrow a phrase from the defunct rock band Timbuk3, the future's so bright that we've got to wear shades.

\section*{Acknowledgements}

I would like to thank Panos Kevrekidis for the invitation to write this article.  I would also like to acknowledge Martin Centurion, Panos Kevrekidis, Alex Nicolin, Yaron Silberberg, and Ian Spielman for reading and providing critical comments on drafts of this manuscript.  Finally, I also thank Markus Oberthaler, Moti Segev, and Yaron Silberberg for permission to use their figures.

%%%%%%%%%%%%%%%%%%%%%%%%%%%%%%%%%%

%\bibliographystyle{plain}
%\bibliography{bec,ref,fpu}

\end{document}